\documentstyle[multicol,prl,aps]{revtex}

\begin{document}
\draft
\title{{\bf On Critical Exponents and the Renormalization of the Coupling Constant} 
\\
{\bf in Growth Models with Surface Diffusion}}
\author{{\sc H.K.\ Janssen}}
\address{Institut f\"{u}r Theoretische Physik III, Heinrich-Heine-Universit\"{a}t,\\
40225 D\"{u}sseldorf, Germany }
\date{\today}
\maketitle

\begin{abstract}
It is shown by the method of renormalized field theory that in contrast to a
statement based on a mathematically ill-defined invariance transformation
and found in most of the recent publications on growth models with surface
diffusion, the coupling constant of these models renormalizes nontrivially.
This implies that the widely accepted supposedly exact scaling exponents are
to be corrected. A two-loop calculation shows that the corrections are small
and these exponents seem to be very good approximations.
\end{abstract}


\begin{multicols}{2}
\narrowtext
There has been great interest in nonequilibrium growth processes of surfaces
and interfaces in the last ten years (for general overviews on the work in
this area of statistical physics and the related literature see the articles
of Halpin-Healy and Zhang \cite{HHZ95}, Krug \cite{Med95,K96} and the book
of Barabasy and Stanley \cite{BSt96}). Recently deposition processes in
which surface diffusion constitutes the dominant relaxation mechanism have
been intensively studied. Besides the simulational work on many different
models, stochastic differential equations for the surface growth with
surface diffusion are proposed and studied by scaling and one-loop
momentum-shell renormalization methods in order to find out the universal
scaling behavior. An apparently mathematically ill-defined invariance
transformation leading to supposedly exact scaling exponents plays a crucial
role in this analytical work. This transformation and its consequences have
been taken over, apparently without careful examination, in most of the
related publications and overviews \cite{HHZ95,BSt96}, and are now widely
accepted in the community of surface-growth physicists.

In this letter I explicitly show that the widely accepted invariance is
indeed wrong and the supposedly exact exponents are to be corrected. I will
use the systematic method of renormalized field theory (see e.g. \cite
{A84,ZJ93,BJW76,DDP78}) and calculate the corrections in an $\varepsilon $%
-expansion to second order.

I consider a system evolving by diffusion of atoms at a surface of (suitably
scaled) height $s\left( {\bf x},t\right) $ over a $d$-dimensional
hyper-plane. Thus the deterministic part of the equation of motion for the
stochastic dynamical variable $s$ has to be of the conserved form 
\begin{eqnarray}
\lambda ^{-1}\frac{\partial s}{\partial t} &=&-\nabla ^2\left[ \tau s+\nabla
^2s+\frac g2\left( \nabla s\right) ^2\right] +\zeta  \nonumber \\
\langle \zeta \left( {\bf x},t\right) \zeta \left( {\bf x}^{\prime
},t^{\prime }\right) \rangle &=&2\lambda ^{-1}\,\left( -\nabla ^2\right)
^a\delta \left( {\bf x}-{\bf x}^{\prime }\right) \delta \left( t-t^{\prime
}\right) .  \label{5}
\end{eqnarray}
Here the parameter $a=1$ in the fully conserved case originally studied by
Sun, Guo, and Grant \cite{SGG89} in their first exploration of non-KPZ
universality classes (with $\tau =0$). If the noise is not conserved (shot
noise) the parameter $a=0$. This case of (\ref{5}) was introduced in the
first papers on ideal molecular beam epitaxy (MBE) \cite{WV90,V91,LDS91}. In
the Langevin equation (\ref{5}) $\lambda $ is a kinetic coefficient and $g$
is the coupling constant. The behavior near the critical point $\tau =0$ is
of primary interest. For $\tau >0$ the equations (\ref{5}) are eventually
driven by renormalization to the linear Edwards-Wilkinson equations both
with nonconserved or conserved noise.

For a generalization of the Galilean invariance (tilt symmetry) of the KPZ
equation \cite{KPZ86} Sun, Guo, and Grant introduced as a symmetry of (\ref
{5}) the coordinate transformation 
\begin{equation}
s\rightarrow s+{\bf a\cdot x},\qquad {\bf x\rightarrow x}-\lambda gt{\bf a}%
\nabla ^2  \label{3}
\end{equation}
which appears to be mathematically ill-defined because it mixes coordinates
and linear differential operators in a coordinate substitution rule (the
questionableness of the rule (\ref{3}) shows up in a definition of the
transformation of a product of functions $f_1\left( {\bf x}\right) f_2\left( 
{\bf x}\right) =f_2\left( {\bf x}\right) f_1\left( {\bf x}\right) $). Based
on the transformation (\ref{3}) the authors state that the coupling constant 
$g$ changes only trivially under renormalization and derive supposedly exact
scaling exponents. The ill-defined transformation (\ref{3}) has been used
for various equations of the type (\ref{5}) to derive critical exponents.
Only few authors express some doubt concerning the use of (\ref{3}) \cite
{V91,TN91}.

In the following I show by an explicit two-loop calculation that the
coupling constant does indeed change non-trivially under renormalization.
Therefore the supposedly exact exponents are to be corrected by a quantity $%
\delta $: 
\begin{equation}
\chi =\frac \varepsilon 3-\delta ,\qquad z=4-\frac \varepsilon 3-2\delta 
\label{4}
\end{equation}
where $\varepsilon =d_c-d$ is the deviation from the upper critical
dimension $d_c.$ The roughening exponent $\chi $ and the dynamic exponent $z$
are defined by the structure function $\langle \left| s\left( {\bf x}%
,t\right) -s\left( {\bf 0},t\right) \right| ^2\rangle \sim \left| {\bf x}%
\right| ^{2\chi }f\left( t/\left| {\bf x}\right| ^z\right) $, where $f$ is a
scaling function. I calculate the correction $\delta $ to order $\varepsilon
^2$ and find that it is always a small but nonzero positive quantity which
has not been detected in simulations up to now.

To develop a renormalized field theory I introduce the dynamic functional 
\cite{DD76,J76,J89,J92,BJW76} corresponding to the Langevin equations (\ref
{5}) 
\begin{eqnarray}
{\cal J} &=&\int d^dx\,dt\,\lambda \left\{ \lambda ^{-1}\widetilde{s}\dot{s}%
+\tau \nabla \widetilde{s}\nabla s%
%TCIMACRO{\TeXButton{rule}{\rule[0.0cm]{0.0cm}{0.4cm}}}
%BeginExpansion
\rule[0.0cm]{0.0cm}{0.4cm}%
%EndExpansion
\right.   \nonumber \\
&&\left. +\nabla ^2\widetilde{s}\left[ \nabla ^2s+\frac g2\left( \nabla
s\right) ^2\right] -\left( \nabla ^a\widetilde{s}\right) ^2\right\} .
\label{6}
\end{eqnarray}
Here $\widetilde{s}$ denotes the response field (the MSR \cite{MSR72}
auxiliary variable). Now all response and correlation functions can be
calculated by path integrals with the weight $\exp \left( -{\cal J}\left[ 
\widetilde{s},s\right] \right) .$ The expansion of the path integrals about
the Gaussian part of ${\cal J}$ generates the (diagrammatic) perturbation
theory which has to be regularized and renormalized because the series are
UV-divergent at the upper critical dimensions that are given by $d_c=4$ for $%
a=0$ and $d_c=2$ for $a=1.$ The propagator $G\left( {\bf q},t\right) =\theta
\left( t\right) \,\exp \left( -\kappa \left( q\right) t\right) $ with $%
\kappa \left( q\right) =\lambda q^2\left( \tau +q^2\right) $ and the
correlator $C\left( {\bf q},t\right) =\chi \left( q\right) \,\exp \left(
-\kappa \left( q\right) \left| t\right| \right) $ with $\chi \left( q\right)
=q^{2\left( a-1\right) }\left( \tau +q^2\right) ^{-1}$ are considered as
functions of the momentum and time. The step function $\theta \left(
t\right) $ is defined with $\theta \left( 0\right) =0.$ The vertex is given
by $V\left( {\bf q}_1;{\bf q}_2,{\bf q}_3\right) ={\bf q}_2\cdot {\bf q}%
_3\,v\left( q_1\right) $ with $v\left( q\right) =-\lambda g\,q^2$. Here $%
{\bf q}_2,{\bf q}_3$ are the ingoing momenta of the two $s$-legs and ${\bf q}%
_1={\bf q}_2+{\bf q}_3$ denotes the outgoing momentum of the $\widetilde{s}$%
-leg. Thus the diagrammatic expansions lead only to one-particle irreducible
vertex functions with (at least) a factor $q^2$ at each (amputated) $%
\widetilde{s}$-leg and a factor ${\bf q}$ at each $s$-leg. A look at the
na\"{\i }ve dimensions of these vertex functions shows that only primitive
divergencies arise which correspond to the terms $\nabla ^2\widetilde{s}%
\nabla ^2s$ and $\nabla ^2\widetilde{s}\left( \nabla s\right) ^2$ in the
dynamic functional (\ref{6}). Therefore the model is renormalizable by the
scheme 
\begin{eqnarray}
s &\rightarrow &\mathaccent"7017{s}=Z^{1/2}s,\quad \widetilde{s}\rightarrow %
\mathaccent"7017{\widetilde{s}}=Z^{-1/2}\widetilde{s},\quad \lambda
\rightarrow \mathaccent"7017{\lambda }=Z\lambda   \nonumber \\
\tau  &\rightarrow &\mathaccent"7017{\tau }=Z^{-1}\tau ,\quad g^2\rightarrow %
\mathaccent"7017{g}^2=A_\varepsilon \mu ^\varepsilon Z^{-3}Z_uu.  \label{10}
\end{eqnarray}
$A_\varepsilon $ =$\left( 4\pi \right) ^{d/2}\Gamma \left( 1+\varepsilon
/2\right) ^{-1}$ is a suitable constant and $\mu $ is the usual inverse
external length scale which makes the renormalized coupling constant $u$
dimensionless. The two renormalization constants $Z$ and $Z_u$ are
determined by minimal renormalization. One easily gets (corresponding to the
known 1-loop results) 
\begin{equation}
Z=1-2^{a-2}\frac u\varepsilon +O\left( u^2\right) ,\quad Z_u=1+O\left(
u^2\right) .  \label{11}
\end{equation}
The incorrect assertion based on the invariance transformation (\ref{3}) is
that $Z_u=1$ to all orders of $u$. In the following I give the derivation of
the second order contribution.

The calculation is greatly simplified due to the following properties of the
diagrammatic contributions. Consider a diagram in the momentum-time picture.
In this representation the restrictions which causality imposes are more
easily exploited than in the momentum-frequency picture; note in particular
the $\theta $-function of the propagator. Due to causality the vertices are
ordered from right to left with increasing time labels. Thus, a diagram with 
$n$ vertices consists of $\left( n-1\right) $ time segments. A Fourier
transformation from time differences to frequencies then generates a
dynamical factor 
\begin{equation}
D\left( \left\{ {\bf q},\omega \right\} \right) =\left[ i\sum_\alpha \omega
_\alpha +\sum_k\kappa \left( {\bf q}_k\right) \right] ^{-1}  \label{12}
\end{equation}
for each time segment. Here $\sum_\alpha \omega _\alpha $ is the sum over
all external frequencies flowing into the segment from the right, and the
set $\left\{ {\bf q}_k\right\} $ are the internal momenta running from right
to left within the segment. In addition to the dynamical factors, each
correlator contributes a factor $\chi \left( {\bf q}\right) $ and each
vertex a factor $V\left( {\bf q}_1;{\bf q}_2,{\bf q}_3\right) $. Finally one
has to integrate over the internal momenta.

I now consider one-particle irreducible diagrams with amputated legs as
contributions to the vertex functions. For the determination of the
UV-divergencies all external momenta and frequencies can be set to zero if
the external momenta coming from the derivatives at the external legs are
included in the amputation procedure. To find the renormalization factor $%
Z_u $, one needs the vertex function which results from the diagrams with
amputated legs corresponding to one $\nabla ^2\widetilde{s}$ and two $\left(
\nabla s\right) $. I now discuss how certain diagrams cancel (this
discussion is analogous to that in \cite{JSch86}). Consider the earliest
vertex of the diagram. If this vertex has an amputated $\left( \nabla
s\right) $-leg, its contributions coming from different permutations of the
propagator (Fig.\ 1a) add to zero because ${\bf q}v\left( -q\right) \chi
\left( q\right) -{\bf q}v\left( q\right) \chi \left( -q\right) =0.$ It is
essential for this cancellation that the vertex considered carries the
earliest time argument, since otherwise the straightforward permutation of
propagators is prohibited by causality. Especially, by this cancellation all
one-loop vertex functions with zero external momenta vanish, and it follows
that the coupling renormalizes only trivially to this order. If the earliest
vertex is an internal one, the contributions coming from the propagator
permutations are added up to the summed vertex (Fig.\ 2a) 
\begin{eqnarray}
S\left( {\bf q}_1,{\bf q}_2,{\bf q}_3\right) &=&\frac{V\left( {\bf q}_1;{\bf %
q}_2,{\bf q}_3\right) }{\chi \left( q_1\right) }  \nonumber \\
&&+\frac{V\left( {\bf q}_2;{\bf q}_3,{\bf q}_1\right) }{\chi \left(
q_2\right) }+\frac{V\left( {\bf q}_3;{\bf q}_1,{\bf q}_2\right) }{\chi
\left( q_3\right) }  \label{13}
\end{eqnarray}
with ${\bf q}_1+{\bf q}_2+{\bf q}_3=0$. The factors $\chi ^{-1}$ result from
the exchange of a propagator line with the corresponding correlator line.

If in the diagrams under consideration the two vertices with the amputated $%
\left( \nabla s\right) $-legs are deleted, all possible diagrams with
internal earliest vertex (the only ones that are nonzero) reduce to one
kernel-diagram (Fig.\ 2b) with a contribution 
\begin{eqnarray}
K\left( {\bf q}_1,{\bf q}_2,{\bf q}_3\right)  &=&\frac{\lambda gq_1^{\,2}}2%
\frac{V\left( {\bf q}_1;{\bf q}_2,{\bf q}_3\right) }{2\kappa \left(
q_1\right) }  \nonumber \\
&&\cdot \frac{\chi \left( q_1\right) \chi \left( q_2\right) \chi \left(
q_3\right) }{\kappa \left( q_1\right) +\kappa \left( q_2\right) +\kappa
\left( q_3\right) }S\left( {\bf q}_1,{\bf q}_2,{\bf q}_3\right)   \label{14}
\end{eqnarray}
which is self-explaining (the first factor $1/2$ results from the double
connection of the two internal vertices). Inserting back the external
vertices allows the discussion of a second cancellation mechanism. The
insertions of an amputated vertex in a time segment with only two internal
lines add to zero if all external momenta are zero (Fig.\ 1b) because ${\bf q%
}v\left( q\right) -{\bf q}v\left( -q\right) =0$. Therefore only the
insertions into the second time segment with three internal lines of the
kernel-diagram (Fig.\ 2b) may lead to a nonvanishing contribution. The
result of the addition of the insertions of one amputated external vertex
with zero external momentum into a three-line segment (Fig.\ 1c) is the
factor 
\begin{equation}
{\bf W}\left( {\bf q}_1,{\bf q}_2,{\bf q}_3\right) =\frac{{\bf q}_1v\left(
q_1\right) +{\bf q}_2v\left( q_2\right) +{\bf q}_3v\left( q_3\right) }{%
\kappa \left( q_1\right) +\kappa \left( q_2\right) +\kappa \left( q_3\right) 
}  \label{15}
\end{equation}
where ${\bf q}_1+{\bf q}_2+{\bf q}_3=0$, and the denominator arises because
a new time segment is produced by an insertion. In the case of the KPZ
equation the functions $v\left( q_i\right) $ are constants and it follows
that these insertions add to zero: ${\bf W}=0$, a result which is easily
generalized to insertions in all segments and has its origin in the Galilean
invariance \cite{JSch86}. But in the models considered here $v\left(
q\right) \sim q^2$, and it is easily seen that ${\bf W}\left( {\bf q}_1,{\bf %
q}_2,{\bf q}_3\right) =0$ only in the case of at least one vanishing
momentum ${\bf q}_i$. In general ${\bf W}\neq 0$ which eventually disproves
all statements on the trivial renormalization of the coupling constant. The
insertions of the two external vertices then lead to the tensor ${\bf %
W\otimes W}$. But after integration over the momenta the integral has to be
a rotational invariant. Thus it must be proportional to the $d$-dimensional
unit tensor and its prefactor can be determined by taking the trace.
Therefore instead of the integral with the tensor ${\bf W\otimes W}$ only
the integral with the trace $\frac 1d{\bf W\cdot W}$ must be calculated to
find the result of the insertion of the two external vertices into the
kernel-diagram. Eventually as the contribution of all two loop diagrams to
the amputated vertex function with zero external momenta and frequencies I
obtain the integrand 
\begin{equation}
I\left( {\bf q}_1,{\bf q}_2,{\bf q}_3\right) =\frac 2dK\left( {\bf q}_1,{\bf %
q}_2,{\bf q}_3\right) {\bf W}\left( {\bf q}_1,{\bf q}_2,{\bf q}_3\right) ^2
\label{16}
\end{equation}
with ${\bf q}_1+{\bf q}_2+{\bf q}_3=0$. The factor $2$ arises from the
permutation of the external amputated $s$-legs.

Performing the integral over the internal loop momenta $R\allowbreak
=\allowbreak \left( 2\pi \right) ^{-2d}\allowbreak \int \prod
d^dq_i\allowbreak \delta \left( {\bf q}_1+{\bf q}_2+{\bf q}_3\right)
\allowbreak I\left( {\bf q}_1,{\bf q}_2,{\bf q}_3\right) $ one observes that
the IR-regulator $\tau >0$ can be neglected in all terms with exception of
the dynamical factor $\left( \kappa \left( q_1\right) +\kappa \left(
q_2\right) +\kappa \left( q_3\right) \right) ^{-3}.$ Then the integration
over the variable $s=q_1^{\,2}+q_2^{\,2}+q_3^{\,2}$ extracts the $%
\varepsilon $-pole, and in the remaining part of the integral the dimension $%
d$ can be safely set to $d_c$. Up to trivial angle integrations the result $%
R $ reduces to a finite two-dimensional integral without any singularities
in the integrand. The elimination of the $\varepsilon $-pole from the
renormalized vertex function finally leads to a definitely nontrivial
renormalization constant 
\begin{eqnarray}
Z_u &=&1-\frac{B_a}\varepsilon u^2+O\left( u^3\right)  \nonumber \\
B_a &=&\left\{ 
\begin{array}{lll}
0.0057406 & \quad \text{if\quad } & a=0 \\ 
0.0167998 & \quad \text{if\quad } & a=1
\end{array}
\right. .  \label{19}
\end{eqnarray}

The RG functions are defined by $\gamma =\left( \partial \ln Z/\partial \ln
\mu \right) _0$, $\gamma _u=\left( \partial \ln Z_u/\partial \ln \mu \right)
_0$, and $\beta =\mu \left( \partial u/\partial \mu \right) _0$ (the index $0
$ denotes differentiation at fixed bare parameters). From the
renormalization of the coupling constant (\ref{10}) one gets $\beta =\left(
-\varepsilon +3\gamma -\gamma _u\right) u$. At the stable nontrivial fixed
point $u_{*}\neq 0$ with $\beta \left( u_{*}\right) =0$ (for $\varepsilon >0$%
), this equation yields the anomalous scaling dimension $\eta =\gamma \left(
u_{*}\right) =\varepsilon /3+2\delta $ of the height-field $s$ where I have
defined $\delta =\gamma _u\left( u_{*}\right) /6$. From the calculated
renormalization factors (\ref{11}) and (\ref{19}), the $\gamma $-functions
are derived as $\gamma =2^{a-2}u+O\left( u^2\right) $ and $\gamma
_u=2B_au^2+O\left( u^3\right) $. Thus from $\beta \left( u_{*}\right) =0$
one finds to first order in the $\varepsilon $-expansion $%
u_{*}=2^{2-a}\varepsilon /3+O\left( \varepsilon ^2\right) $. Hence the
correction is given by 
\begin{eqnarray}
\delta  &=&\frac{4^{2-a}}{27}B_a\,\varepsilon ^2+O\left( \varepsilon
^3\right)   \nonumber \\
&=&\left\{ 
\begin{array}{lll}
0.01361\,\left( \varepsilon /2\right) ^2 & \quad \text{if\quad } & a=0 \\ 
0.002489\,\varepsilon ^2 & \quad \text{if\quad } & a=1
\end{array}
\right. \quad +O\left( \varepsilon ^3\right) .  \label{23}
\end{eqnarray}
For the physical case of a two-dimensional surface with nonconserved noise ($%
a=0,$ $d_c=4,$ $\varepsilon =2$) or a one-dimensional surface with conserved
noise ($a=1,$ $d_c=2,$ $\varepsilon =1$) the numerical prefactors represent
the results for $\delta .$ As I have stated above, these corrections are
always small but nonzero positive numbers. From the anomalous dimension one
finds the roughening exponent $\chi =\left( \varepsilon -\eta \right) /2$
and the dynamic exponent $z=4-\eta $ as anticipated in (\ref{4}). Of course
the scaling law $z-2\chi =4-\varepsilon $ holds, being an exact consequence
of the trivial renormalization of the noise. But the relation $4=z+\chi
=4+\left( \varepsilon -3\eta \right) /2=4-3\delta $, where the first
equality comes from the ill-defined symmetry (\ref{3}), is definitely wrong.

In summary I have shown that in contrast to a statement found in most of the
recent publications on growth models with surface diffusion, the coupling
constant of these models renormalizes non-trivially. This implies that the
widely accepted scaling exponents are to be corrected. But my two-loop
result also shows that these corrections are very small as long as one
accepts the numerical result at least as a qualitative one with respect to
the order of magnitude. Thus the supposedly exact exponents seem to be a
very good approximation and the deviations are not seen in simulations and
experiments up to now. I suppose that the origin of the smallness of the
correction is found in the gradient coupling leading to many (but
incomplete) cancellations between diagrams as well as within the internal
momentum integrals. Therefore I expect the mode-coupling approach which
neglects vertex corrections completely to be appropriate here. I will come
back to this in the future.

I remark that the calculation presented in this letter is the first two-loop
calculation for the vertex functions of the stochastic models considered
here. Concerning these models the work of Das Sarma and Kotlyar \cite{DSK94}
contains a collection of two-loop self-energy diagrams. However these
diagrams are irrelevant for the problem of the second-order renormalization
of the coupling constant. Thus the authors cite the one-loop result $\chi
+z=4$ for the model with nonconserved noise. Based on the ill-defined transformation (\ref{3}) they believe this scaling law to hold to all orders in perturbation theory. As I have shown this statement is incorrect.

\acknowledgments I thank S.\ Theiss for a critical reading of the
manuscript. This work has been supported in part by the SFB 237 of the
Deutsche Forschungsgemeinschaft.

%TCIMACRO{
%\TeXButton{Capt1}{\begin{figure}
%\caption{Cancellation of diagrams: a) an earliest external vertex,
% b) an external vertex inserted in a two-line segment,
% but c) the insertions of an external vertex in a three-line segment
% in all possible ways do not cancel in general,
% (time increases always from right to left, the arrows indicate the time direction
% of the propagators). }
%\label{fig1}
%\end{figure}}}
%BeginExpansion
\begin{figure}
\caption{Cancellation of diagrams: a) an earliest external vertex,
 b) an external vertex inserted in a two-line segment,
 but c) the insertions of an external vertex in a three-line segment
 in all possible ways do not cancel in general,
 (time increases always from right to left, the arrows indicate the time direction
 of the propagators). }
\label{fig1}
\end{figure}%
%EndExpansion

%TCIMACRO{
%\TeXButton{Capt2}{\begin{figure}
%\caption{The fundamental diagram: 
%a) an internal earliest vertex with all its propagator permutations,
% b) the two-loop kernel diagram,
% (time and momenta flow from right to left).}
%\label{fig2}
%\end{figure}}}
%BeginExpansion
\begin{figure}
\caption{The fundamental diagram: 
a) an internal earliest vertex with all its propagator permutations,
 b) the two-loop kernel diagram,
 (time and momenta flow from right to left).}
\label{fig2}
\end{figure}%
%EndExpansion

\end{multicols}

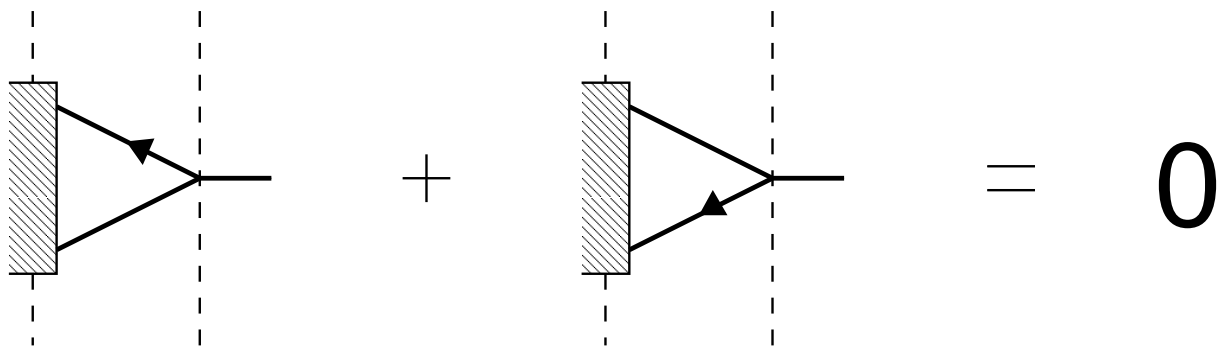
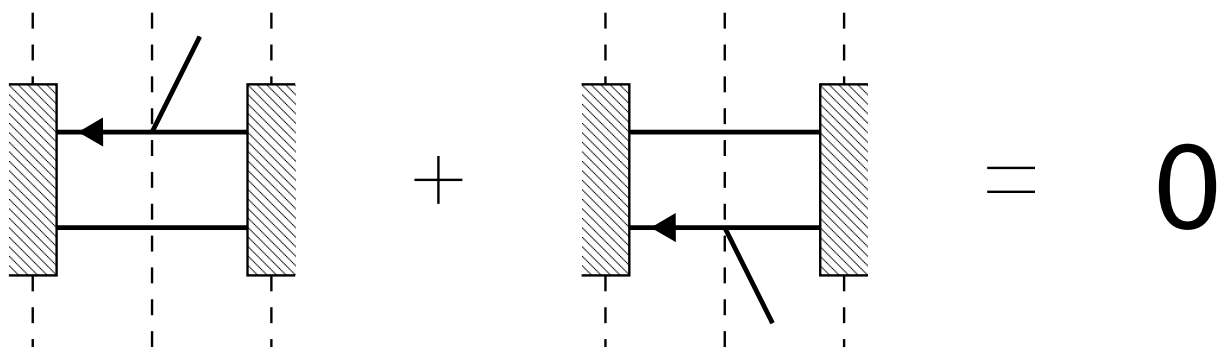
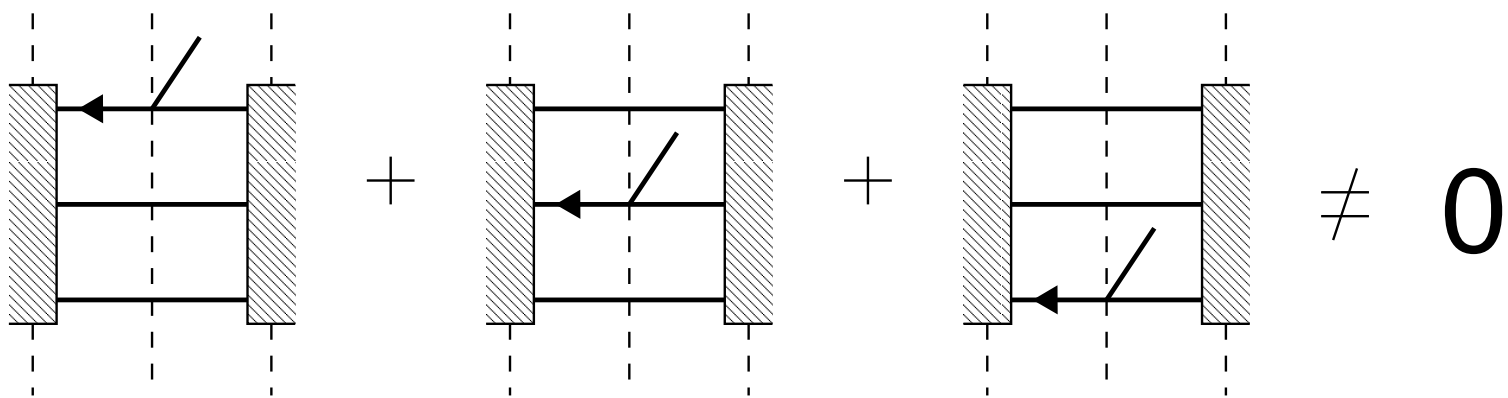
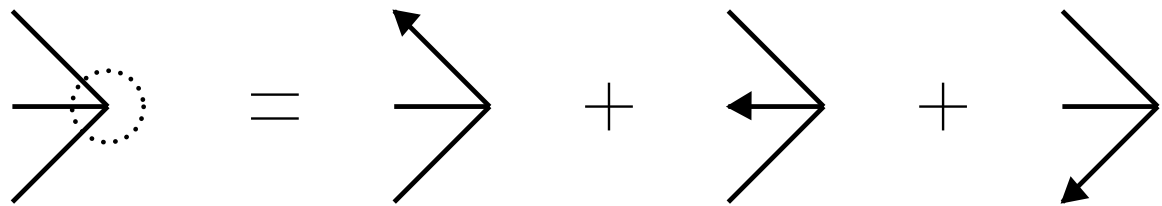
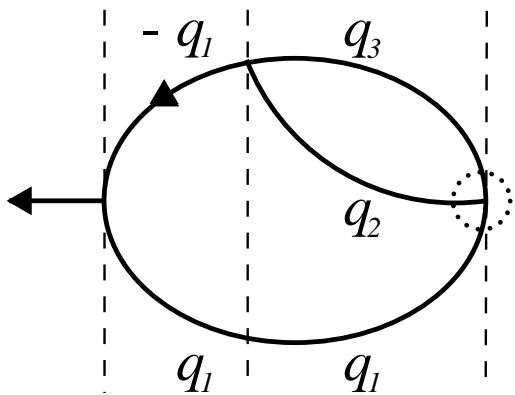
\end{document}